# Managing Self-Phase Modulation in Pseudo-linear Multimodal and Monomodal Systems

M. Zitelli, M. Ferraro, F. Mangini, and S. Wabnitz, *Senior Member, IEEE*

*Abstract*—We propose a new semi-analytical model, describing the bandwidth evolution of pulses propagating in dispersion managed (DM) transmission systems using multimodal graded-index fibers (GRIN) with parabolic index. The model also applies to monomodal fiber DM systems, representing the limit case where beam self-imaging vanishes. The model is successfully compared with the direct integration of the (1+1)D nonlinear Schrödinger equation for parabolic GRIN fibers, and to experimental results performed by using the transmission of femtosecond pulses over a 5 m span of GRIN fiber.
At the high pulse powers that are possible in multimodal fibers, the pulse bandwidth variations produced by the interplay of cumulated dispersion and self-phase modulation can become the most detrimental effect, if not properly managed. The analytical model, numerical and experimental results all point to the existence of an optimal amount of chromatic dispersion, that must be provided to the input pulse, for obtaining a periodic evolution of its bandwidth. Results are promising for the generation of spatio-temporal DM solitons in parabolic GRIN fibers, where the stable, periodic time-bandwidth behaviour that was already observed in monomodal systems is added to the characteristic spatial beam self-imaging.

*Index Terms*—Fiber nonlinear optics, optical solitons, optical fibers.

## I. INTRODUCTION

DISPERSION managed (DM) fiber optic transmission systems have been widely investigated over the past twenty years, mainly because of the enhanced power and moderate time jitter properties of the transmitted signal. A dispersion managed system is composed by a repetition of two opposite dispersion fiber segments, composing a span of cumulative length $L_{span}$; dispersion and losses are compensated at each span, according to different design geometries. The propagating pulse experiences periodical variations of pulse width, chirp and bandwidth, owing to the interplay of the fibers' dispersion and Kerr nonlinearity. If the alternating fibers in each span have chromatic dispersion $\beta_{2\_1}$ and $\beta_{2\_2}$ (ps$^2$/km) and length $L_1$ and $L_2$, the DM strength is defined as $S = (L_1|\beta_{2\_1}| + L_2|\beta_{2\_2}|)/T_{FWHM}^2$, with $T_{FWHM}$ the minimum pulse width. In order to have stable periodical behavior for the transmitted Gaussian pulse, the map strength must overcome a critical value, and the amount of cumulated dispersion at the end of each span must have a small anomalous or zero (negligible mean dispersion) value, respectively. An optical amplifier is inserted at the end of each compensated span, in order to recover the fiber losses.

In [1], the authors investigated numerically, by means of the nonlinear Schrödinger equation (NLSE), the formation of monomodal stable soliton-like pulses in dispersion compensated maps composed by anomalous-normal-anomalous dispersion monomodal fiber segments, with a resulting span length of 200 km. Simulations launched 20 ps FWHM Gaussian pulses with 650 μW peak power. It was observed that: (i) the path-averaged dispersion must be anomalous, of the order of -0.1 ps$^2$/km; (ii) the period of dispersion compensation must be short with respect to the nonlinear length; (iii) the local dispersion must differ significantly from zero in order to avoid dispersive wave generation.

In Ref. [2], an empirical scaling law was proposed for the optimal launch power of pulses in monomodal DM systems, including losses and periodical amplifiers. The law indicated the need for an enhanced power with respect to traditional soliton systems, with a dispersion equal to the average dispersion of the DM system. When the amplification and the dispersion period are comparable, it was necessary to modify the position at which a transform-limited pulse source should be located within the dispersion map.

Several works used the NLSE, the Gaussian pulse ansatz, and the variational approach to find a set of differential equations describing the evolution of the pulse width and chirp of a propagating pulse in a monomodal DM system, without [3] and with lumped amplifiers and fiber losses [4][5].

Other works brought to similar sets of differential equations describing the evolution of parameters such as γ, related with the spectral width, and the chirp C [6], or the root-mean square (rms) pulse width and the integral pulse chirp, through the rms momentum equations method [7]. It was shown that steady

Submitted on October 12, 2020.
This work was supported by: the European Research Council (ERC) under the European Union's Horizon 2020 research and innovation programme (No. 874596, No. 740355). The Italian Ministry of University and Research (R18SPB8227).
M. Zitelli, M. Ferraro, S. Wabnitz are with the Department of Information Engineering, Electronics and Telecommunications (DIET), Sapienza University of Rome, Via Eudossiana 18, 00184 Rome, Italy (e-mail: mario.zitelli@uniroma1.it).
F. Mangini is with the Department of Information Engineering (DII), University of Brescia, Via Branze 38, 25123 Brescia, Italy.

pulse propagation is possible, whenever the average dispersion lies in the anomalous domain.

The variational approach was also used in the case of single-mode fiber links, in order to provide a physical description of the pulse bandwidth and chirp at the fibers' boundaries and mid-points; and to find a critical map strength $S$ for optimal transmission [8].

In [9], optimal dispersion management techniques were provided for the minimization of collision-induced frequency shifts in wavelength division multiplexed systems (WDM). Once again, differential equations describing the pulse width and chirp were derived from the NLSE, by using the Lagrangian density variation method; equations were then extended to the case of colliding pulses.

In [10], the dynamical behavior of single-channel transmission in monomodal standard fibers with strong dispersion management and linear compensating devices was theoretically and numerically analyzed. Single-pulse propagation was compared with single-channel 40 Gbit/s transmissions, to highlight the relevant roles played by nonlinearity-induced spectrum distortion and intra-channel pulse interactions.

In most fiber systems, the Kerr nonlinearity or self-phase modulation (SPM) produces chirp and bandwidth changes on isolated pulses affected by cumulated dispersion. Considering a single anomalous fiber span with dispersion $\beta_2$ (ps$^2$/km), nonlinear coefficient $\gamma$ (1/W/km), with propagating pulse width $T_0 = T_{FWHM}/1.763$, the traditional local solitonic condition is reached when the dispersion length $L_D = T_0^2/|\beta_2|$ equals the nonlinear length $L_{NL} = 1/\gamma P_0$, being $P_0$ the pulse peak power.

In DM systems, the local dispersion length $L_D$ is considerably shorter than the local nonlinearity length $L_{NL}$; the DM solitonic regime is different from the NLSE or conventional soliton regime: it is reached when the temporal breathing pulse experiences SPM, that compensates for a moderate amount of anomalous residual dispersion, and mostly for large duty cycle signal formats. This regime shows optimum transmission for a anomalous mean dispersion of the dispersion map. The so-called pseudo-linear regime in DM systems is characterized by a local dispersion length which is much shorter than the nonlinearity length, but it also exhibits optimum transmission at zero net residual dispersion. This regime is mostly relevant for low duty-cycle signal formats. For moderate values of fiber dispersion (for example, for D≥4 ps/nm/km), the solitonic regime exhibits reduced performances with respect to the pseudo-linear regime [11].

Besides SPM affecting single pulse transmission, channel performance is also affected by pulse-to-pulse nonlinear effects, specifically, intra-channel cross-phase modulation (IXPM) and intra-channel four-wave mixing (IFWM).

IXPM arises from the interaction of adjacent pulses at same wavelength [12]; it generates a frequency shift on each pulse, that depends on the presence of neighbouring marks or spaces and, through dispersion, it leads to timing jitter generation [13].

IFWM results from four-wave mixing between different spectral components of dispersed overlapping pulses; after full dispersion compensation, it causes the appearance of shadow pulses in correspondence of the spaces, and amplitude jitter for the marks. IFWM dominates with respect to IXPM for relatively high pulse powers.

Both IXPM and IFWM are mitigated by the use of symmetric dispersion maps [14], characterized by a normal pre-compensation which is equal to half the cumulated dispersion of the anomalous fiber span, and opposite in sign.

When the pulse power reaches high values, as in the case of DM systems using multimodal fibers or monomodal large modal area fibers, SPM is responsible for bandwidth changes which may extend up to a significant fraction of the pulse bandwidth itself. Whenever SPM causes a pulse bandwidth increase, performance impairments are observed after the optical filter at the receiver side; on the other hand, if SPM causes a bandwidth decrease, the pulse cannot recover the initial pulse width and interferes with the adjacent bit time slots, thus causing inter-symbol interference (ISI). In this regime, SPM may become the most detrimental nonlinear effect, unless it is managed by a proper choice of the pre-compensation.

All of the previous mentioned studies considered monomodal DM fiber systems. Few attempts have been made to extend the DM technique to two-dimensional spatio-temporal breathers in planar waveguides [15][16]. On the other hand, multimodal graded-index fibers (GRIN) are interesting candidates for DM transmission systems, when considering the high pulse powers that can be injected in the fiber and, consequently, the improved optical signal-to-noise ratio and system capacity. Dispersion compensation after a GRIN fiber span may be achieved, for example, by using lumped devices, such as fiber Bragg gratings, followed by lumped amplifiers.

In the next section, we introduce a different, with respect to the variational method, semi-analytical approach which is capable of describing the bandwidth evolution of pulses in a DM system using multimodal GRIN fibers with parabolic index. Our method is based on analytically solving the NLSE over discrete fiber steps $\Delta z$, in order to find a semi-analytical expression that calculates the pulse bandwidth evolution when affected by cumulated dispersion, Kerr nonlinearity and periodical intensity oscillations due to beam spatial self-imaging (SSI) in a parabolic GRIN fiber [17]. The method also applies to monomodal fiber DM systems, for the particular condition of negligible SSI.

In section 3 the model will be compared with the direct numerical simulation of the (1+1)D NLSE for GRIN multimodal systems and for monomodal systems; in section 4, the bandwidth evolution of femtosecond pulses will be experimentally investigated on a 5 m span of GRIN parabolic fiber, and compared with model predictions, showing good agreement.

II. THEORY

A Gaussian optical pulse propagating in a DM system suffers dispersion-induced pulse width variations at each fiber span, as well as bandwidth variations induced by SPM; this behavior is observed for input pulse peak powers lower than the local solitonic threshold for anomalous dispersion fibers. At the system input, a pre-compensation $\beta_{2pre}$ ($s^2$) is used in order to convert those variations into periodic oscillations. In other words, by properly choosing the value of $\beta_{2pre}$, the output pulse will periodically recover its input bandwidth, thus preserving the transmission quality. As a matter of fact, the pseudo-linearity condition does not prevent the pulse from suffering important bandwidth changes in propagation regions where the minimum pulse width is recovered, thus imposing the use of optimal pre-compensation techniques.

From the variational theory applied to the nonlinear term of the NLSE [18][19][20], optical transmission in a parabolic graded-index (GRIN) multimodal fiber, as well as in a monomodal step-index (SI) fiber, can be described by a single nonlinear Schrödinger equation (NLSE), which include the second order dispersion term $\beta_2$, third order dispersion $\beta_3$ ($s^3/m$), Kerr nonlinearity $\gamma$ and linear loss $\alpha$

$$\frac{\partial A(z,t)}{\partial z} = -i\frac{\beta_2}{2}\frac{\partial^2 A}{\partial t^2} + \frac{\beta_3}{6}\frac{\partial^3 A}{\partial t^3} - \frac{\alpha}{2}A + i\gamma(z)|A|^2 A \quad , \tag{2.1}$$

with $A(z,t)$ the (1+1)D field complex amplitude. For a multimodal parabolic GRIN fiber with core radius $r_c$, core index $n_{co}$, nonlinear index $n_2$ ($m^2/W$), and relative index difference $\Delta$, the nonlinear coefficient $\gamma$ ($m^{-1}W^{-1}$) is periodical in $z$ with

$$\gamma(z) = \frac{n_2 \omega_0}{c\pi w^2(z)} \quad , \tag{2.2}$$

$$w(z) = w(0)\left[\cos^2\left(\pi\frac{z}{z_p}\right) + C \cdot \sin^2\left(\pi\frac{z}{z_p}\right)\right]^{1/2}, \tag{2.3}$$

with $w(z)$ the beam waist, that undergoes periodical self-imaging with period $z_p = \pi r_c/\sqrt{2\Delta}$; the ratio between minimum and maximum beam area is given by

$$C = \frac{4c^2 z_p^2}{[\pi n_{co}\omega_o w^2(0)]^2} \quad . \tag{2.4}$$

For $C = 1$, the beam waist is a constant, and Eq. 2.1 describes monomodal transmission, both for parabolic GRIN fibers and SI fibers.

A semi-analytical formula to describe the pulse bandwidth evolution in pseudo-linear systems can be obtained by applying the split-step method to Eq. 2.1 over a distance $\Delta z$; the formula is much faster than the direct integration of eq. 2.1, and can be used for both monomodal and parabolic GRIN fiber systems to calculate the pulse bandwidth evolution, and to quickly search for the optimal pre-compensation which causes periodic band behavior.

We consider the transmission of a Gaussian pulse with input pulse width $T_{FWHM}$, and $T_0 = T_{FWHM}/(2(ln2)^{1/2})$, and with amplitude $A_0$. After a distance $z$, the pulse acquires a cumulated dispersion $\beta_{2tot}(z)$ ($s^2$), including pre-compensation, fiber dispersion and any possible subsequent dispersion compensation: its complex amplitude can be written as [21]

$$A(z,t) = \frac{A_0 T_0}{[T_0^2 - i\beta_{2tot}(z)]^{1/2}} exp\left[-\frac{t^2}{2[T_0^2 - i\beta_{2tot}(z)]}\right]; \tag{2.5}$$

Therefore, the pulse power reads as

$$|A(z,t)|^2 = \frac{A_0^2}{[1+C_{disp}^2]^{1/2}} exp\left[-\frac{t^2}{T_0^2(1+C_{disp}^2)}\right] \quad , \tag{2.6}$$

with $C_{disp}(z) = \beta_{2tot}(z)/T_0^2$ the cumulated dispersive chirp. At fiber distance $z$, SPM is responsible for pulse nonlinear phase shift $\Delta\Phi_{NL}(z,t)$ over a distance interval $\Delta z$; by considering that the pulse peak power at distance $z$ is affected by losses and dispersion-induced pulse broadening, the nonlinear distance $L_{NL} = 1/\gamma(z)|A|^2$ is given by

$$L_{NL}(z) = \frac{[1+C_{disp}^2]^{1/2}}{\gamma(z)A_0^2 exp(-\alpha z)} \quad . \tag{2.7}$$

The nonlinear phase shift over $\Delta z$ is $\Delta\phi_{NL}(z,t) = |A(z,t)|^2 \Delta z_{eff}/[A_0^2 L_{NL}(z)]$, with $\Delta z_{eff} = (1 - exp(-\alpha\Delta z))/\alpha$, when accounting for losses. The corresponding pulse bandwidth variation is

$$\Delta\omega(z,t) = -\frac{\partial \Delta\phi_{NL}}{\partial t} = \frac{T_0}{a}\left[-\frac{2t}{a^2}exp\left(-\frac{t^2}{a^2}\right)\right]\frac{\Delta z_{eff}}{L_{NL}(z)} \quad , \tag{2.8}$$

with $a = T_0\left(1 + C_{disp}^2(z)\right)^{1/2}$; the bandwidth change must be calculated at the instants of maximum and minimum spectral width, i.e., for $t = \pm a/\sqrt{2}$, obtaining

$$\Delta\omega_{max}(z) = const \cdot 0.858 \cdot \frac{\Delta z_{eff}}{T_0[1+C_{disp}^2(z)]L_{NL}(z)} \cdot$$
$$sign(C_{disp}(z)) \quad . \tag{2.9}$$

In Eq. 2.9, the *sign*() operator is needed, in order to account for positive (negative) bandwidth changes whenever the overall cumulated dispersion is normal (anomalous). *const* is a scaling factor accounting for the ratio between the FWHM bandwidth $\Delta\omega_{max}$ and the bandwidth change at $t = \pm a/\sqrt{2}$, to be calculated via the comparison with numerical simulations of Eq. 2.1; it was found $const \cdot 0.858 = 1.7$. The nonlinear length accounts for fiber losses and, in the case of GRIN fibers, for beam self-imaging; the term $1 + C_{disp}^2$ takes into account the dispersion-induced pulse broadening. Let us point out that cumulated dispersion includes pre-compensation, fiber cumulated dispersion, and periodic dispersion compensation. After each compensated span, the pulse power must be recovered to its initial value by using optical amplifiers.

Eq. 2.9 provides a measure of the interplay between SPM and cumulated dispersion, and it shows that pulse bandwidth changes may be experienced by both pre-compensated and non-precompensated pulses. In fact, these changes occur when SPM acts on the pulse while it is cumulating a net dispersion, and at the same time, it still possesses significant peak power.

By summing the bandwidth changes in Eq. 2.9 for Δz sufficiently small, and over a number of compensated fiber spans, it is possible to describe the pulse bandwidth evolution vs. distance at a given pre-compensation value. It is also possible to search for an optimal pre-compensation, that will produce a periodical variation of the pulse bandwidth along the distance z. In the case of monomodal fibers, Δz can be as large as 1 m, while for parabolic GRIN fibers it must be reduced to 50 μm or less, in order to account for the short spatial scale of periodic beam self-imaging.

For pulse powers up to 20% of the local solitonic threshold, the SPM-induced bandwidth changes may represent a considerable fraction of the initial bandwidth; if not properly managed, these bandwidth changes will produce significant impairments at the system output.

If no pre-compensation is used, Eq. 2.9 predicts monotonic bandwidth changes at each fiber span. In the case of anomalous dispersion fibers, such as standard monomodal fibers at telecom wavelengths (SMF), the pulse bandwidth suffers repeated bandwidth reductions after each DM span, that prevent the initial pulse width from recovering. This is shown by the red curve in Fig.1, illustrating the bandwidth evolution of a 0.83 ps pulse in 5 spans, composed by anomalous dispersion fiber followed by a linear dispersion compensator and an amplifier. Whenever full pre-compensation is used ( $\beta_{2pre} = -\beta_2 L_{span}$), the pulse bandwidth experiences repeated increases (see green curve in Fig.1), thus introducing channel cross-talk and filtering impairments. On the other, an optimal pre-compensation value leads to a periodical evolution of the bandwidth, and stable transmission (see the blue curve in Fig.1).

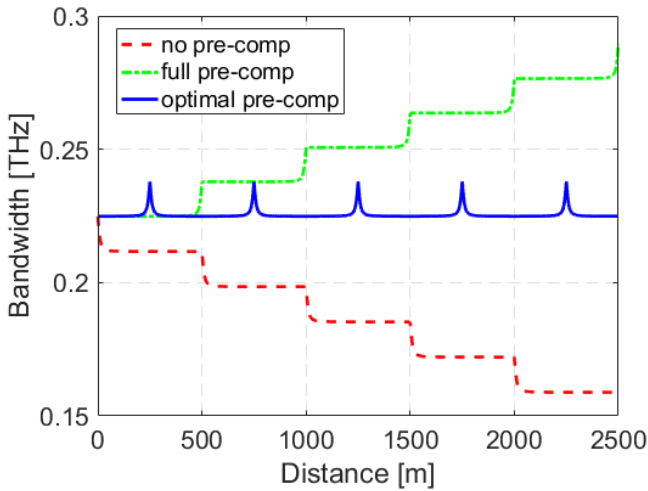

Fig. 1. Example of pulse bandwidth evolution in a pseudo-linear dispersion compensated system with anomalous dispersion fiber and: no pre-compensation, full pre-compensation, or optimal pre-compensation.

III. NUMERICAL SIMULATIONS

The theory of section 2 has been compared to direct numerical simulations of Eqs. 2.1 to 2.3, solved by means of the well-known split-step Fourier method [22]. The simulated system is a sequence of identical spans, composed by a transmission fiber and a linear dispersion compensator, followed by an optical amplifier that recovers entirely fiber losses. At the system input, the pulse is pre-chirped by an amount of pre-compensation, or input chirp $\beta_{2pre}$, with opposite sign with respect to that of the transmission fiber. The two cases of SMF and parabolic GRIN fiber were considered, for pulse durations which are typical of high-speed communication systems.

*A. Parabolic GRIN fiber, multimodal dispersion-managed systems*

DM systems employing parabolic GRIN fibers need to overcome the impairments caused by modal dispersion; this is minimized at the 850 nm wavelength, therefore in the normal dispersion regime; dispersion compensation can be achieved, for example, by using lumped devices, such as fiber Bragg gratings.

The model of Eqs. 2.7 to 2.9 needs to consider a periodically variable nonlinear coefficient $\gamma(z)$, and the sum of bandwidth changes must be performed with a step size $\Delta z \leq 50$ μm, in order to properly describe the beam self-imaging process.

Numerical simulations were also performed by using Eqs. 2.1-2.3, with $n_2 = 2.7 \times 10^{-20}$ m²/W, modal waist $w(0) = 15$ μm at fiber input, dispersion $\beta_2 = 36.2$ ps²/km at 850 nm, loss $\alpha = 2.21$ dB/km, and neglecting third-order dispersion $\beta_3$; the core radius was $r_c = 25$ μm, the core index $n_{co} = 1.465$, and the relative index difference $\Delta = 0.0103$.

After each span, the accumulated dispersion was compensated by an amount $-\beta_2 L_{span}$ by using ideal linear lumped dispersion compensators, and losses were recovered by means of ideal lumped amplifiers.

We simulated two cases, with Gaussian pulse durations of $T_{FWHM} = 0.83$ and $3.33$ ps, corresponding to systems with a baud rate of 400 and 100 GB/s, respectively. Pulses were launched at the 850 nm wavelength, with a peak power of 17.98 and 1.12 W, respectively. Although in the normal dispersion region no solitonic effect is observed, the power levels were chosen for a fair comparison with the monomodal case which is considered in the next section.

Figs. 2a and 2b illustrate the variation of the output-to-input bandwidth ratio as a function of pre-compensation, for 0.83 and 3.33 ps pulse widths, respectively; solid lines are obtained from the semi-analytical theory of Eqs. 2.9, 2.2, and 2.3, while dots indicate the results of numerical simulations from Eqs. 2.1 to 2.3. As can be seen, our theory shows good agreement with numerical simulations for multimodal fiber transmission, indicating that, for longer pulses, an optimal pre-compensation value is necessary for minimizing the SPM-induced impairments. For shorter pulses with intermediate levels of pre-compensation, the bandwidth evolution is faster, thus limiting the net bandwidth change. The required pre-compensation is anomalous, before entering the normal dispersion fiber; as a consequence, the pulse bandwidth evolution with distance exhibits periodical spikes with an opposite sign with respect to the case shown in Fig. 1; this is better illustrated in Fig. 3, showing the bandwidth evolution of a 0.833 ps pulse at 18 W peak power in 5 × 500 m of GRIN fiber at the 850 nm wavelength, with the optimal pre-compensation of -9 ps². In Fig.3 we reported the bandwidth evolution described by the model of eq. 2.9, and compared it to the numerical simulation of Eq. 2.1.

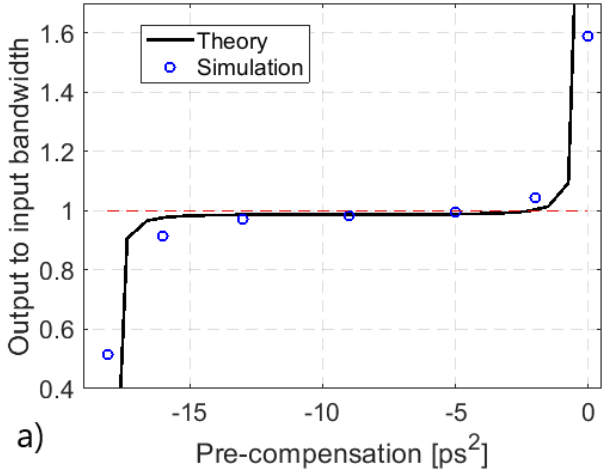

a)

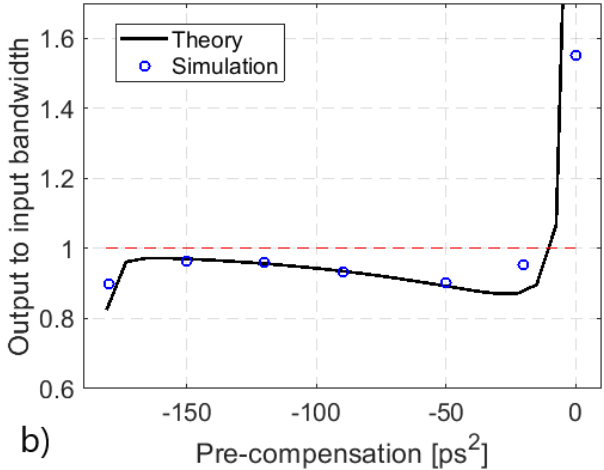

b)

Fig. 2. Ratio of output to input pulse bandwidth, in the cases of: (a) 0.83 ps and 5 x 500 m of GRIN fiber, (b) 3.33 ps and 5 x 5 km GRIN fiber spans.

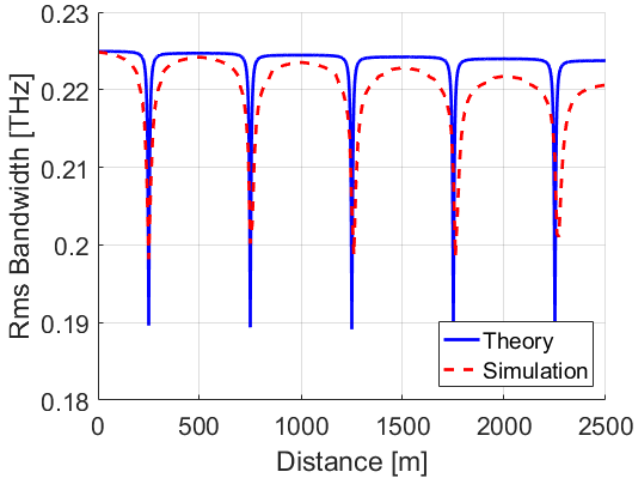

Fig. 3 – Bandwidth evolution for a 0.833 ps pulse, 18 W peak power, 850 nm, pre-compensation of -9 ps² in 5 × 500 m of GRIN fiber; comparison of model prediction and numerical simulation of Eq. 2.1.

### B. Monomodal dispersion managed systems

DM systems employing monomodal fibers at wavelengths around 1550 nm are commonly used in high-speed telecommunications. These systems experience minimum transmission losses, and operate in the anomalous dispersion region: among them, the case of dispersion-compensated SMF fiber spans is widely used. We tested four cases with Gaussian pulse durations of $T_{FWHM} = 0.83, 3.33, 6.66, 13.33$ ps, corresponding to systems operating at the baud rate of 400, 100, 50, and 25 GB/s, respectively. Pulses were launched at the 1550 nm wavelength, with a peak power of 17.98, 1.12, 0.28, and 0.07 W respectively, corresponding to 20% of the power for reaching the monomodal soliton regime in an SMF fiber, neglecting the losses. Even for lower powers, we obtained nearly constant optimal pre-compensation values, which shows that in the pseudo-linear regime the optimal $\beta_{2pre}$ has only a weak dependence on the launch power. Whereas for higher peak power values in the anomalous dispersion regime, the fundamental soliton behavior starts to be effective, giving rise to a different propagation regime.

SMF spans are described by Eq. 2.1 with constant $\gamma$, $n_2 = 2.7 \times 10^{-20}$ m²/W, modal waist $w = 5.04$ μm, effective area $A_{eff} = 80$ μm², dispersion $\beta_2 = -20$ ps²/km, loss $\alpha = 0.2$ dB/km; we neglected the third-order dispersion $\beta_3$. After each fiber span of length $L_{span}$, a dispersion compensation amount of $-\beta_2 L_{span}$ was applied, and losses were fully recovered by means of ideal amplifiers.

Figures 4a to 4d illustrated the dependence on the amount of pre-compensation of the ratio of pulse output to input bandwidth, in the cases of: a): 0.83 ps and 5 spans of 5 km of SMF fiber; b): 3.33 ps and 5 × 20 km spans; c): 6.66 ps and 5 × 50 km spans; d): 13.33 ps and 5 × 75 km spans. The fiber length was chosen as the minimum value, which is necessary to obtain a complete periodical evolution of the pulse bandwidth; for longer $L_{span}$ values, the bandwidth evolution does not change, and the same optimal pre-compensation values were found. Solid lines in Fig. 4 are obtained from the semi-analytical theory of Eq. 2.9, while dots represent numerical simulations from Eq. 2.1.

As shown by Fig. 4a, for an ultra-short pulse with $T_{FWHM} = 0.83$ ps, there is a large range of pre-compensation values where the output pulse bandwidth equals the input. It is also clear that the absence of pre-compensation, as well the full pre-compensation, provides bandwidth changes of the order of 60% for the considered case of pulsewidth and power.

Figs. 4b and 4c, obtained for $T_{FWHM} = 3.33$ and 6.66 ps, respectively, show the existence of an optimal pre-compensation value, which is consistent with both theory and numerical simulations. Let us recall that the optimal pre-compensation values are still valid for larger values of $L_{span}$, because in pseudo-linear systems bandwidth changes are limited in the fiber portion of minimum pulsewidth, and are negligible in the remaining portions.

Finally, Fig. 4d obtained for $T_{FWHM} = 13.33$ ps, shows that larger differences appear between theory and simulations, as we move away from the pseudo-linear transmission conditions.

As already observed, shorter pulses breathe in time more rapidly; the interplay between SPM and cumulated dispersion produces negligible effects at any value of used pre-compensation, when the cumulated dispersion passes rapidly from anomalous to normal (or the opposite). Conversely, the same interplay causes larger net bandwidth changes when the cumulated dispersion does not change its sign (if pre-compensation is null or it is total). On the contrary, longer pulses suffer considerable bandwidth net changes when a wrong pre-compensation is used, because the interaction between SPM and dispersion acts over longer distances.

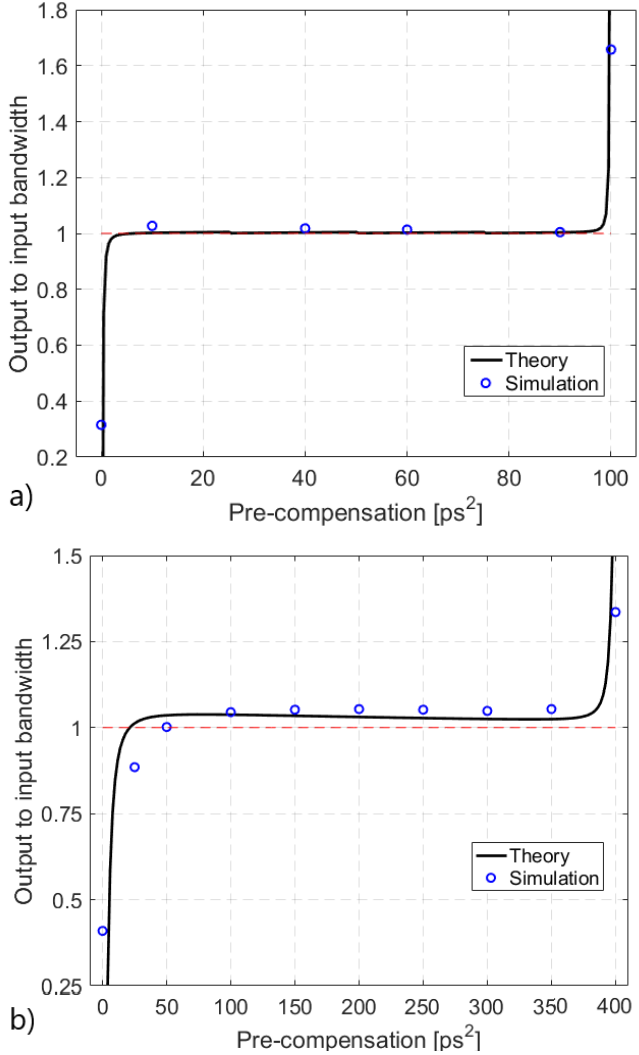

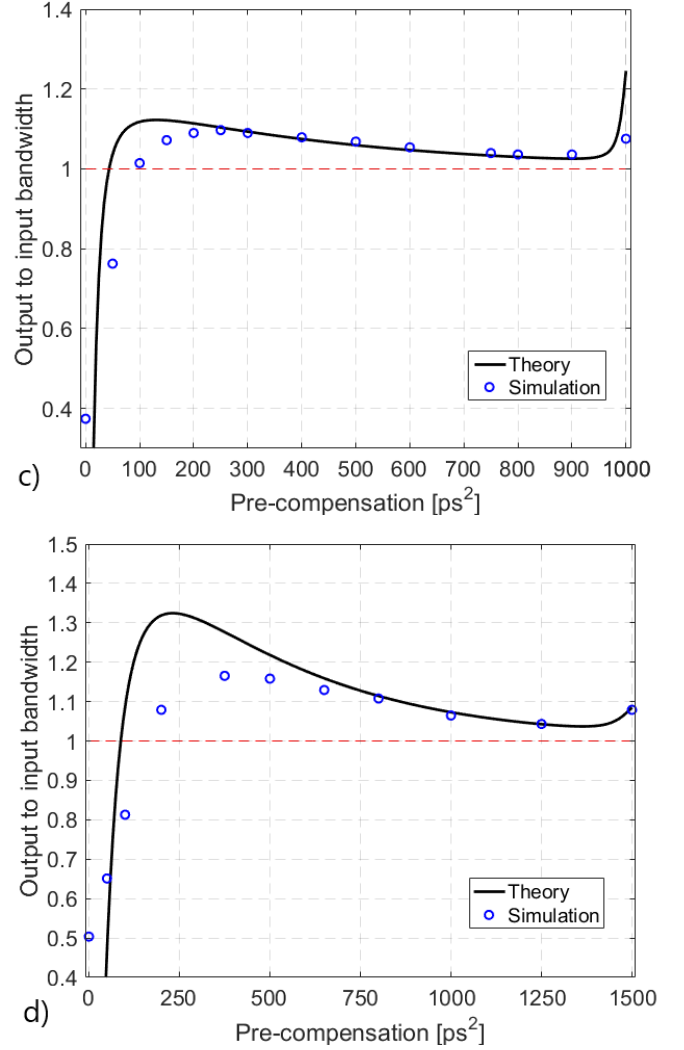

Fig. 4. Ratio of the output to input pulse bandwidth vs. input pre-compensation, in the cases of: a): 0.83 ps and 5 × 5 km of SMF, b): 3.33 ps and 5 × 20 km SMF, c): 6.66 ps and 5 × 50 km SMF, d): 13.33 ps and 5 × 75 km SMF spans.

## IV. EXPERIMENTAL RESULTS WITH PARABOLIC GRIN FIBER

The experimental setup used for the study of pulse bandwidth evolution as a function of pre-compensation in a parabolic GRIN fiber consisted of a femtosecond Yb-based laser (Light Conversion Pharos PH1-SP 10W), generating variable temporal duration pulses at 1030 nm, with 100 kHz repetition rate, and Gaussian beam shape ($M^2 = 1.3$). The laser is equipped with internal pulse-stretched amplification; by properly tuning the internal compression factor, it was possible to generate pulse widths varying between 180 fs and 8 ps, corresponding to a pre-compensation $\beta_{2pre}$ between -0.3 ps$^2$ and 0.4 ps$^2$; laser pulse width and shape were measured by using a FROG autocorrelator (APE pulseCheck). The beam was focused by a 50 mm lens into a 5 m span of parabolic GRIN fiber, with input $1/e^2$ diameter of approximately 20 µm. Pulse energy was adjusted using an external attenuator.

GRIN fiber has a core radius $r_c = 25$ µm, cladding radius 62.5µm, cladding index $n_{clad} = 1.45$, relative index difference Δ= 0.0103, length 5 m, and dispersion $\beta_2 = 18.9$ $ps^2$/km at 1030 nm.

At the fiber output, a micro-lens followed by a second lens focused the beam to an optical spectrum analyzer (OSA, Yokogawa AQ6370D) and a power meter (Gentec XLP12-3S-VP-INT-D0). The output near field was also projected on a 1550 nm camera (Gentech Beamage 4M-IR) in order to check the correct coupling and modal distribution.

In order to stress the model predictions for high pulse powers, the pulse energy was adjusted to either 0.096 nJ or 0.96 nJ at the fiber input; this corresponded to pulses with either 500 W or 5 kW peak power, when transform-limited to a 180 fs pulsewidth. Pulse energy was kept constant while broadening the input pulse, by adding pre-compensation.

Figure 5 shows the measured pulse bandwidth (dots) vs. input pulse pre-compensation, for the two input pulse energy levels. Solid curves are the predictions from the semi-analytical theory of Eq. 2.9; for anomalous input pre-compensation $\beta_{2pre} = -\beta_2 L_{span}$=-0.095 $ps^2$, our model predicts a maximum reduction of the output pulse bandwidth, while in the absence of pre-compensation (for $\beta_{2pre} = 0$ $ps^2$) a maximum increase is expected. For intermediate pre-compensation values, the pulse bandwidth suffers opposite sign variations along with the fiber, obtaining an output rms bandwidth closer to the input value of 1.04 THz.

The model curve at 5 kW maximum peak power fits well the experimental data for normal pre-compensation values. However, for anomalous dispersion values, the model of Eq. 2.9 fails in predicting band reductions up to zero or even negative, which is obviously an unphysical result, and differs from the moderate band reduction which is observed experimentally. The model at 500 W peak power fits well to experimental data for normal pre-compensation, and provides a better approximation, with respect to the 5 kW case, for anomalous pre-compensation values. The differences between theory and experiments for anomalous values of pre-compensation may be attributed to pulse distortions at low bandwidth values, which are not accounted by our theoretical model.

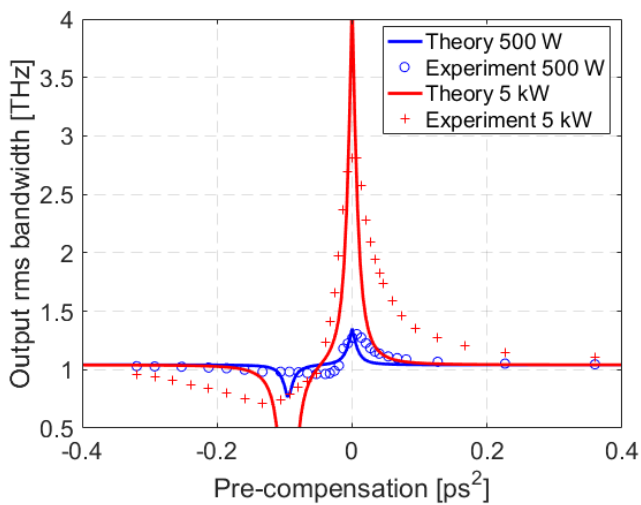

Fig. 5. Pulse output bandwidth vs. input pre-compensation in 5 m of parabolic GRIN fiber. Dots: experiment; solid curves: model of eq. 2.9. Input peak power is 500W or 5 kW for a minimum pulsewidth of 180 fs at 1030 nm.

## V. CONCLUSIONS

The use of parabolic GRIN fibers in DM systems offers interesting options to increase the signal pulse power well beyond the limits of monomodal systems, therefore increasing the signal-to-noise ratio and channel capacity. Here we demonstrate by means of a new semi-analytical formula, the direct integration of the (1+1)D NLSE, and by experimental results, that parabolic GRIN fibers are able to support the techniques of bandwidth and dispersion management which have already been developed for monomodal transmission systems. Parabolic GRIN fibers are potentially able to support spatio-temporal DM solitons, which constitute a new transmission regime with properties of stability and time-bandwidth periodicity similar to what already seen in the monomodal case, but with the further property of stable beam self-imaging acting on a very short spatial period, of the order of 0.5 mm. Furthermore, spatio-temporal solitons offer the unique property to temporally trap in them the several spatial modes that may carry a beam in GRIN fibers [19][10].

The proposed semi-analytical model, describing the bandwidth evolution of propagating pulses, was derived by analytically integrating the NLSE with a periodically varying nonlinear coefficient $\gamma(z)$, and it was compared to both the direct integration of the (1+1)D nonlinear Schrödinger equation for parabolic GRIN fibers, and to experimental results performed at the 1030 nm wavelength with femtosecond pulses, propagating in a 5 m span of GRIN fiber. It was shown that an optimal pre-compensation dispersion is needed, in order to obtain a periodical behavior of the pulse bandwidth, which provides the first basic step, which is necessary for a stable propagation of spatio-temporal DM solitons.

**Mario Zitelli** obtained the M.Sc. in Electronics Engineering and the Ph.D. in Applied Electromagnetism from Sapienza University of Rome in 1995 and 1999 respectively. From 1997 to 1999 he was with Ugo Bordoni Foundation (FUB); since 1999 he has worked as a senior researcher and R&D manager in several telecom companies: Pirelli Submarine Telecom Systems (Milan, Italy), QPlus Networks Inc. (Long Beach, CA), Santel Networks Inc. (Fremont, CA), CNET France-Telecom (Lannion, France), TelCon (Rome, Italy), OCEM Airfield Technology (Valsamoggia, Italy), Leonardo (Pomezia, Rome), Selex System Integrations (Overland Park, Kansas, USA), Argos Ingegneria (Rome, Italy), Photoneco ltd (London, UK). Since July 2019 he is researcher in Telecommunications at Sapienza University of Rome. His research activities include nonlinear fiber transmission, spatio-temporal nonlinear effects, advanced optical modulation formats, photometry. He is author/co-author of over 40 international refereed papers and conference presentations, 5 international patents, and was member/co-director of 5 research projects from EU, NATO, FILAS.

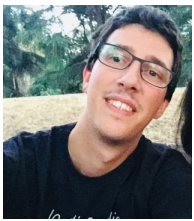

**Mario Ferraro** received the Bachelor's degree in Physics "cum laude and special mention to the curriculum" in 2013 from University of Calabria (Italy) and the Master degree in Physics "cum laude" in 2015 from "La Sapienza" University of Rome (Italy). He has been awarded as excellent graduated student of "La Sapienza" University of Rome in 2016 and he received the Ph.D. in Physics in 2019 from the University of Côte d'Azur in Nice (France). His research activity has concerned nonlinear optics in nanodisordered media as well as semiconductor-based hyperbolic metameterials. Since January 2020 he is a postdoctoral fellow at the Engineering department of "La Sapienza" University of Rome working on ultrafast nonlinear phenomena in multimode optical fibers.

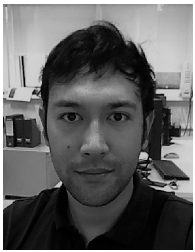

**Fabio Mangini** received his B.Sc. in Clinical Engineering and M.Sc. in Biomedical Engineering "cum laude" from "La Sapienza" University of Rome, Italy in 2005 and 2008, respectively. He earned his Ph.D. in Electromagnetism from the Department of Information Engineering, Electronics, and Telecommunications of the same University in 2014. In June 2014 and in May 2015, he won the "Young Scientist Award" from URSI (International Union of Radio Science). In January 2017, he won the "Ph.D. ITalents" prize and in October 2018 he won the "Marabelli prize". Between 2009 and 2015 he worked at the Laboratory of Electromagnetic Fields II at "La Sapienza" University of Rome. His research activities focus on guiding structures, numerical methods, theoretical scattering models, optical propagation, anisotropic media, metamaterials, biomedical applications, and cultural-heritage applications. Since April 2019, he has been working, as a researcher, at the Electromagnetic Fields and Photonics group at the University of Brescia.

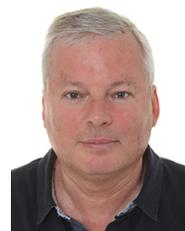

**Stefan Wabnitz** obtained the Laurea Degree in Electronics Engineering from Sapienza University of Rome in 1982, the MS in Electrical Engineering from Caltech in 1983, and the PhD in Applied Electromagnetism from the Italian Ministry of Education in 1988.

He was with the Ugo Bordoni Foundation between 1985 and 1996. From 1996 until 2007 he was full professor at the University of Burgundy in Dijon, France. Between 1999 and 2003 he was with Alcatel Research and Innovation Labs in France, and with Xtera Communications in Allen, Texas, USA. Since 2007 until 2018 he was full professor in Electromagnetic Fields at the University of Brescia, Italy. Since November 2018 he is full professor in Telecommunications at Sapienza University of Rome. His research activities involve nonlinear propagation effects in optical communications and information processing devices.

Prof. Wabnitz is the author and co-author of over 700 international refereed papers, conference presentations, and book chapters. He is the Editor-in-Chief of Elsevier's Optical Fiber Technology, a Fellow member of the Optical Society of America, and senior member of IEEE-Photonics Society.